\documentclass[conference]{IEEEtran}
\usepackage{cite}
\usepackage{amsmath,amssymb,amsfonts}
\usepackage{algorithmic}
\usepackage{graphicx}
\usepackage{textcomp}
\usepackage{xcolor}
\usepackage{url}
\usepackage{hyperref}
\usepackage{booktabs}
\usepackage{multirow}

\def\BibTeX{{\rm B\kern-.05em{\sc i\kern-.025em b}\kern-.08em
    T\kern-.1667em\lower.7ex\hbox{E}\kern-.125emX}}

\begin{document}

\title{Intent-Driven Dynamic Chunking: Segmenting Documents to Reflect Predicted Information Needs}

\author{\IEEEauthorblockN{Christos Koutsiaris}
\IEEEauthorblockA{\textit{Dept. of Computer Science \& Information Systems} \\
\textit{University of Limerick}, Ireland \\
24220094@studentmail.ul.ie}
\IEEEauthorblockA{\textit{Cloud ERP - UX Foundation} \\
\textit{SAP P\&E}\\
christos.koutsiaris@sap.com}
}

\maketitle

\begin{abstract}
Breaking long documents into smaller segments is a fundamental challenge in information retrieval. Whether for search engines, question-answering systems, or retrieval-augmented generation (RAG), effective segmentation determines how well systems can locate and return relevant information. However, traditional methods, such as fixed-length or coherence-based segmentation, ignore user intent, leading to chunks that split answers or contain irrelevant noise. We introduce Intent-Driven Dynamic Chunking (IDC), a novel approach that uses predicted user queries to guide document segmentation. IDC leverages a Large Language Model to generate likely user intents for a document and then employs a dynamic programming algorithm to find the globally optimal chunk boundaries. This represents a novel application of DP to intent-aware segmentation that avoids greedy pitfalls. We evaluated IDC on six diverse question-answering datasets, including news articles, Wikipedia, academic papers, and technical documentation. IDC outperformed traditional chunking strategies on five datasets, improving top-1 retrieval accuracy by 5\% to 67\%, and matched the best baseline on the sixth. Additionally, IDC produced 40--60\% fewer chunks than baseline methods while achieving 93--100\% answer coverage. These results demonstrate that aligning document structure with anticipated information needs significantly boosts retrieval performance, particularly for long and heterogeneous documents.

\vspace{0.3em}
\noindent\textbf{Code:} \url{https://github.com/unseen1980/IDC}
\end{abstract}


\section{Introduction}

Breaking long documents into well-chosen smaller segments is a fundamental preprocessing step in information retrieval systems. From search engines and question-answering applications to retrieval-augmented generation (RAG), documents must be split so that each segment can be efficiently indexed, retrieved, and presented to users or downstream models. In practice, nearly every modern retrieval system performs some form of document chunking. However, \textit{how} these chunks are defined can greatly influence performance; even small changes in chunking strategy can noticeably affect retrieval recall and precision. Despite this impact, many implementations treat chunking as a simplistic, ad-hoc procedure rather than as a core algorithmic component informed by end-user needs.

The most common approach, fixed-length chunking, divides text into uniform blocks (e.g., every 200 tokens). While simple, this method is arbitrary: it often cuts through sentences or logical topics, separating context from content. If the window is too small, a single answer can be fragmented across multiple chunks; if too large, each chunk may contain extraneous text that dilutes relevant information. Fixed segmentation is also highly sensitive to the chosen segment length; if not tuned carefully, retrieval quality drops markedly.

Coherence-based methods (e.g., TextTiling~\cite{hearst1997texttiling}, C99~\cite{choi2000advances}) improve on this by respecting discourse boundaries, keeping related ideas together. However, they remain \textbf{query-agnostic}: they optimize for internal document structure rather than the user's information need. A coherent section might still be too broad for a specific query, or an answer might span two coherent sections. This misalignment between document segments and user queries leads to suboptimal retrieval: relevant information may be buried in irrelevant text or fragmented across multiple chunks.

Existing solutions like document expansion (e.g., docT5query~\cite{nogueira2019doctttttquery}) address vocabulary mismatch by adding predicted queries to text, but they do not alter the underlying segmentation. A retrieval system might still return chunks that contain answers mixed with unrelated content, simply because the document was segmented without regard to specific questions.

We propose \textbf{Intent-Driven Dynamic Chunking (IDC)}, a method that realigns document segmentation with user intent. IDC first predicts a set of likely user queries (intents) for a document using a generative model. It then employs a dynamic programming algorithm to segment the text such that each chunk optimally answers one of these predicted questions. By making segmentation intent-aware, IDC ensures that chunks are ``answer-sized'' and focused, containing complete, relevant information without excessive noise.

The motivation for IDC arose from real industrial challenges. In developing a semantic search system for SAP's Fiori technical documentation, we observed that basic chunking strategies held the system back. Engineers seeking specific answers (e.g., ``How do I use API X?'' or ``What does error code Y mean?'') often had to sift through multiple irrelevant chunks or piece together fragmented information. This disconnect between how documents were segmented and the questions users asked made search inefficient. IDC addresses this gap by anticipating user questions during segmentation.

The key contributions of this work are:
\begin{itemize}
    \item We introduce IDC, a novel algorithm that adapts document segmentation to predicted user intents using dynamic programming optimization.
    \item We evaluate IDC on six QA benchmarks across four domains, showing that it improves Recall@1 on five datasets (with gains from 5\% to 67\%) and ties the best baseline on the sixth.
    \item We demonstrate that IDC produces 40--60\% fewer chunks than baselines while achieving higher answer coverage (93--100\%), making it efficient for indexing.
    \item We analyze the efficiency and cost of IDC, showing it adds minimal overhead suitable for offline indexing ($<$\$0.01 per long document).
\end{itemize}

\section{Related Work}

\subsection{Document Segmentation Methods}

Document segmentation research spans several decades. Fixed-length chunking remains common due to simplicity, but early work showed its limitations: Callan~\cite{callan1994passage} found that fixed windows often divide answers between chunks. Wartena~\cite{wartena2013segmentation} confirmed that retrieval performance ``breaks down'' quickly when segment length deviates from optimal values.

Coherence-based approaches emerged to address these issues. TextTiling~\cite{hearst1997texttiling} detects topic shifts by analyzing lexical cohesion, placing boundaries at ``valleys'' of low similarity. C99~\cite{choi2000advances} clusters sentences by semantic similarity to identify topic boundaries. Barzilay and Lapata~\cite{barzilay2008modeling} introduced entity-based coherence modeling for discourse understanding. More recently, Koshorek et al.~\cite{koshorek2018text} framed segmentation as supervised learning with neural models, and Ghinassi et al.~\cite{ghinassi2024recent} surveyed transformer-driven segmentation advances that leverage deep contextual embeddings.

While coherence-based methods produce internally consistent segments, they remain query-agnostic, optimizing for document structure without considering what users might ask. This motivates our intent-driven approach.

\subsection{Query-Aware Document Expansion}

Research in query-aware retrieval has largely focused on document expansion. The doc2query method~\cite{nogueira2019document} predicts likely questions a document can answer and appends them to the text before indexing, bridging vocabulary gaps. Its successor docT5query~\cite{nogueira2019doctttttquery} used the T5 transformer to generate more diverse, fluent questions with improved retrieval gains.

Subsequent work extended this paradigm. InPars~\cite{bonifacio2022inpars} used GPT-3 to create synthetic query-document pairs as training data for retrievers. Promptagator~\cite{dai2022promptagator} demonstrated that prompting large language models can yield useful query variations with minimal examples.

However, these expansion methods do not alter document segmentation. The added queries become part of the text in each document's index entry, but the underlying splitting remains unchanged. If important information is split across chunks due to suboptimal segmentation, appending questions cannot fix that fragmentation. IDC extends the intuition of query prediction from expansion to \textit{structure}, using predicted queries not just to enrich content, but to drive how the document is segmented.

\section{Methodology}

\subsection{Overview of IDC}

Intent-Driven Dynamic Chunking realigns document segmentation with user information needs through two main offline stages: (1)~\textit{Intent Simulation}, where likely user queries are predicted for the document, and (2)~\textit{Boundary Optimization}, where the document is segmented to maximize alignment between chunks and these predicted intents.

\subsection{Intent Simulation}

We generate a set of hypothetical user intents $Q = \{q_1, q_2, \ldots, q_M\}$ for document $D$ using Gemini 2.5 Flash. The LLM is prompted to generate questions the document can answer, covering its main topics and key details. To ensure topic coverage, we employ section-wise generation for longer documents and use stochastic decoding (top-$k$ sampling) for diversity.

The number of generated intents adapts to document complexity: short documents ($<$100 sentences) receive 10--15 questions, while long documents ($>$400 sentences) receive 35--40 questions. This adaptive strategy ensures adequate coverage without over-segmentation. After generation, we filter redundant questions by computing cosine similarity between their embeddings; if two questions exceed a similarity threshold (0.85), we retain only one.

\subsection{Sentence Embedding and Scoring}

The document is split into $N$ sentences $S = \{s_1, s_2, \ldots, s_N\}$. Both sentences and predicted intents are encoded into a shared vector space using a transformer-based sentence embedding model (1536-dimensional embeddings). For a candidate chunk $C_{i,j}$ spanning sentences $i$ to $j$, the chunk embedding is computed as the average of its constituent sentence embeddings. The \textit{intent relevance} score is:
\begin{equation}
R(C_{i,j}) = \max_{q \in Q} \cos(\mathbf{e}(C_{i,j}), \mathbf{e}(q))
\end{equation}
where $\mathbf{e}(\cdot)$ denotes the embedding function. $R(C_{i,j})$ quantifies how well the chunk could answer at least one predicted question.

\subsection{Boundary Optimization}

We find segmentation $\mathcal{S} = \{C_1, C_2, \ldots, C_k\}$ that maximizes the utility function:
\begin{equation}
U(\mathcal{S}) = \sum_{m=1}^k R(C_m) - \lambda \sum_{m=1}^k |C_m|^2 - \beta(k-1)
\end{equation}
where $\lambda$ is a length penalty (discouraging overly long chunks) and $\beta$ is a boundary penalty (discouraging over-segmentation). Because $|C_m|^2$ grows quickly with chunk size, $\lambda$ is typically very small (e.g., 0.0005 after tuning) to allow context-rich chunks without excessive penalty.

We solve this efficiently using dynamic programming. Let $f(j)$ be the maximum utility for optimally segmenting sentences 1 through $j$. The recurrence is:
\begin{equation}
f(j) = \max_{0 \le i < j} \{ f(i) + R(C_{i+1,j}) - \lambda |C_{i+1,j}|^2 - \beta \}
\end{equation}
with $f(0) = 0$. We only consider chunks within a maximum length $L$ (e.g., 10--15 sentences), reducing complexity to $O(N \times L)$, which is essentially linear in document length.

After the DP solution, we apply light post-processing: merging very short adjacent chunks with the same intent, and splitting overly long chunks at natural paragraph boundaries if needed.

\textbf{Design Rationale:} We deliberately chose a hybrid architecture that uses the LLM for semantic reasoning (intent prediction) while employing dynamic programming for structural optimization (boundary selection). An alternative approach, prompting an LLM to directly insert segment markers, would suffer from several limitations: (1) LLMs generate text left-to-right, making greedy local decisions that cannot guarantee globally optimal segmentation; (2) LLM outputs are difficult to control precisely, whereas DP allows explicit hyperparameter tuning (our ablation studies showed that tuning $\lambda$ improved R@1 by 8.5\%); (3) asking an LLM to segment a long document risks hallucination, sentence omission, or content alteration, whereas DP operates on embeddings and preserves document integrity; and (4) DP segmentation runs in milliseconds, while LLM-based segmentation would require generating thousands of output tokens. This hybrid design leverages each component's strengths: LLMs for semantic understanding, algorithms for structural optimization.

\section{Experimental Setup}

\subsection{Datasets}

We evaluated IDC on six question-answering datasets spanning four domains (Table~\ref{tab:datasets}): news articles (NewsQA), Wikipedia (SQuAD), academic papers (arXiv, Qasper), and technical documentation (Fiori). These datasets vary in length (12--495 sentences) and structure, providing a comprehensive evaluation across document types.

\begin{table}[h]
\centering
\caption{Dataset characteristics}
\label{tab:datasets}
\small
\begin{tabular}{lccc}
\toprule
\textbf{Dataset} & \textbf{Domain} & \textbf{Docs} & \textbf{QA Pairs} \\
\midrule
NewsQA & News & 1 & 15 \\
SQuAD 1-doc & Wikipedia & 1 & 12 \\
SQuAD 2-doc & Wikipedia & 2 & 293 \\
arXiv & Academic & 1 & 15 \\
Qasper & Academic & 10 & 10 \\
Fiori & Technical & 1 & 15 \\
\bottomrule
\end{tabular}
\end{table}

\subsection{Baselines}

We compared IDC against four baseline segmentation strategies:
\begin{itemize}
    \item \textbf{Fixed-Length}: Non-overlapping 6-sentence chunks
    \item \textbf{Sliding Window}: 6-sentence chunks with 50\% overlap
    \item \textbf{Coherence-Based}: TextTiling-like topic boundary detection
    \item \textbf{Paragraph-Based}: Natural paragraph breaks as boundaries
\end{itemize}
All methods used identical preprocessing (sentence tokenization), embedding models, and hybrid retrieval (60\% dense + 40\% BM25).

\subsection{Evaluation Metrics}

We used Recall@1 (R@1), Recall@5 (R@5), and Mean Reciprocal Rank (MRR). R@1 measures the fraction of queries where the top-ranked chunk contains the answer, which is critical for QA systems. We also report chunk counts and answer coverage (percentage of answers fully contained within single chunks).

\section{Results}

\subsection{Retrieval Performance}

Table~\ref{tab:results} presents the main retrieval results. IDC achieved the highest R@1 on five of six datasets and tied on the sixth (Qasper).

\begin{table}[h]
\centering
\caption{Retrieval Performance (Recall@1, Recall@5, MRR)}
\label{tab:results}
\small
\begin{tabular}{lccc}
\toprule
\textbf{Dataset / Method} & \textbf{R@1} & \textbf{R@5} & \textbf{MRR} \\
\midrule
\multicolumn{4}{l}{\textit{NewsQA}} \\
\quad IDC & \textbf{0.933} & \textbf{1.000} & \textbf{0.956} \\
\quad Best Baseline & 0.867 & 0.867 & 0.867 \\
\midrule
\multicolumn{4}{l}{\textit{SQuAD 1-doc}} \\
\quad IDC & \textbf{0.917} & \textbf{1.000} & \textbf{0.958} \\
\quad Best Baseline & 0.917 & 0.917 & 0.917 \\
\midrule
\multicolumn{4}{l}{\textit{arXiv (495 sentences)}} \\
\quad IDC & \textbf{0.667} & \textbf{0.933} & \textbf{0.789} \\
\quad Best Baseline & 0.400 & 0.800 & 0.530 \\
\midrule
\multicolumn{4}{l}{\textit{Fiori}} \\
\quad IDC & \textbf{0.533} & \textbf{0.933} & \textbf{0.686} \\
\quad Best Baseline & 0.333 & 0.733 & 0.502 \\
\midrule
\multicolumn{4}{l}{\textit{SQuAD 2-doc (n=293)}} \\
\quad IDC & \textbf{0.689} & \textbf{0.952} & \textbf{0.793} \\
\quad Best Baseline & 0.655 & 0.951 & 0.752 \\
\midrule
\multicolumn{4}{l}{\textit{Qasper}} \\
\quad IDC & 0.250 & 0.500 & 0.333 \\
\quad Best Baseline & \textbf{0.250} & \textbf{0.600} & \textbf{0.367} \\
\bottomrule
\end{tabular}
\end{table}

\begin{figure}[t]
\centering
\includegraphics[width=\columnwidth]{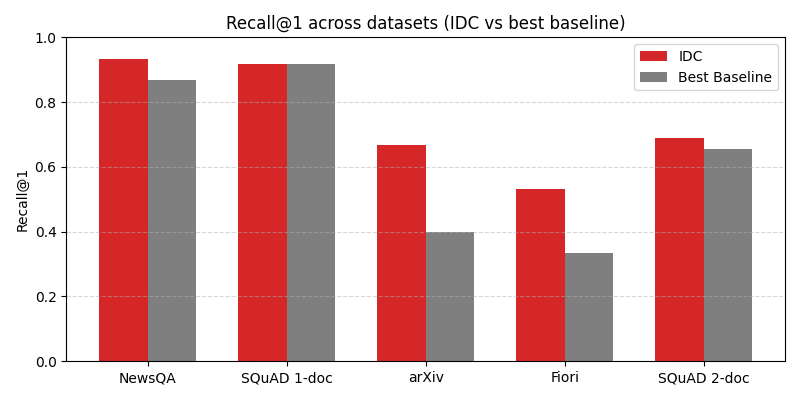}
\caption{Recall@1 across datasets. IDC (red) consistently matches or exceeds the best baseline (gray), with largest gains on long documents (arXiv +67\%, Fiori +60\%).}
\label{fig:recall1}
\end{figure}

\begin{figure}[t]
\centering
\includegraphics[width=\columnwidth]{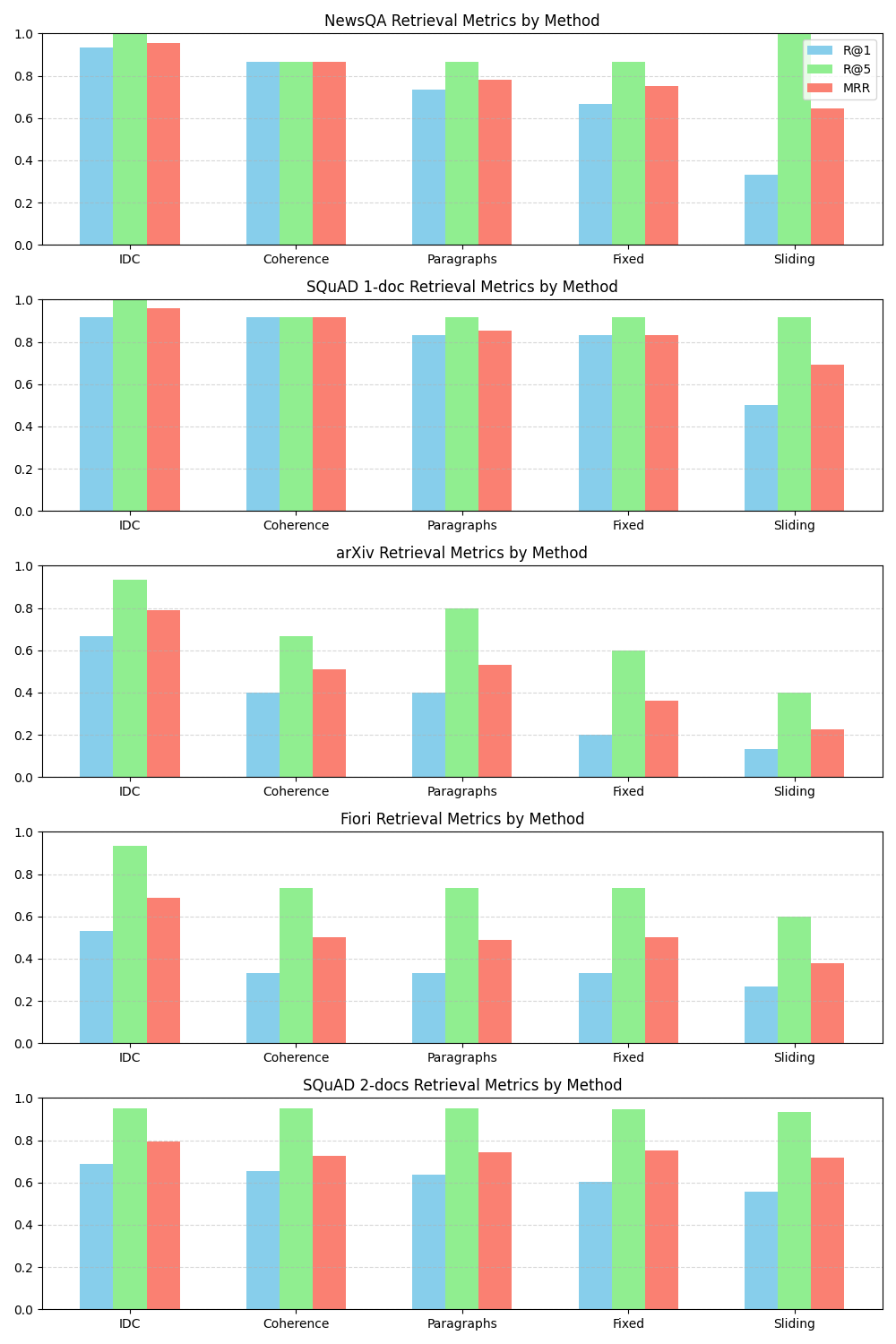}
\caption{Complete retrieval metrics (R@1, R@5, MRR) across all datasets and methods. IDC achieves the highest or tied-highest scores on 5 of 6 datasets.}
\label{fig:all_metrics}
\end{figure}

IDC's improvements were most pronounced on long, heterogeneous documents. On the 495-sentence arXiv paper, IDC achieved R@1 of 0.667 versus 0.400 for baselines, a \textbf{67\% relative improvement}. On Fiori technical documentation, IDC reached 0.533 versus 0.333 (\textbf{+60\%}). On the large SQuAD 2-doc dataset (293 queries), IDC's improvement was statistically significant ($p<0.05$, Cohen's $d\approx0.41$).

The Qasper dataset was an exception: IDC tied with the Paragraph baseline on R@1 (0.250) but showed slightly lower R@5 (0.500 vs 0.600) and MRR (0.333 vs 0.367). This suggests that for highly structured academic papers where each section naturally aligns with specific questions, paragraph-based segmentation can be equally effective. The structured nature of research papers, with clear section boundaries corresponding to distinct topics, provides natural ``intent alignment'' that IDC cannot significantly improve upon.

\subsection{Segmentation Efficiency}

\begin{figure}[t]
\centering
\includegraphics[width=\columnwidth]{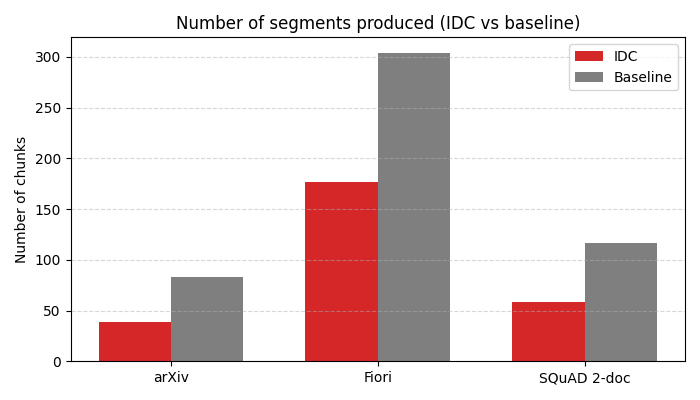}
\caption{Number of chunks produced by IDC vs baselines. IDC generates 40--60\% fewer chunks while achieving higher retrieval performance.}
\label{fig:chunks}
\end{figure}

\begin{figure}[t]
\centering
\includegraphics[width=\columnwidth]{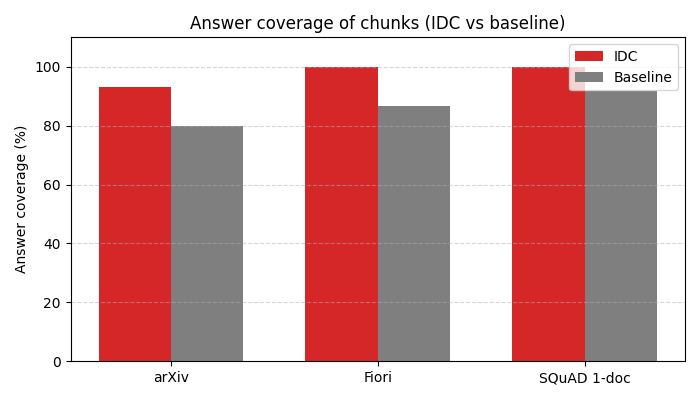}
\caption{Answer coverage: percentage of questions whose answer is fully contained within a single chunk. IDC achieves 93--100\% coverage, compared to 80--87\% for baselines.}
\label{fig:coverage}
\end{figure}

IDC produced significantly fewer chunks than baselines while achieving better retrieval (Figure~\ref{fig:chunks}). On arXiv, IDC created 39 chunks versus 83 for Fixed-length (53\% reduction). On Fiori, IDC produced 177 chunks versus 304 for Fixed (42\% reduction). Fewer chunks means smaller index sizes and faster retrieval.

Despite fewer chunks, IDC achieved higher answer coverage (Figure~\ref{fig:coverage}). On arXiv, IDC covered 93.3\% of answers within single chunks, compared to 80\% for Fixed. On Fiori, IDC achieved 100\% coverage versus 86.7\% for baselines. This demonstrates that IDC's intent-guided boundaries place cuts more intelligently, keeping complete answers intact.

\subsection{Efficiency Analysis}

\textbf{Offline Preprocessing:} IDC takes 1--2 seconds per short document and 10--15 seconds for very long documents ($>$400 sentences). Intent generation dominates this cost ($\sim$1s via Gemini 2.5 Flash API), while DP segmentation is fast ($<$200ms).

\textbf{Query Latency:} Online retrieval is identical for IDC and baselines ($\sim$500ms, dominated by query embedding and index lookup). IDC's preprocessing is entirely offline.

\textbf{Cost Analysis:} Using Gemini 2.5 Flash pricing, costs vary by document length:
\begin{itemize}
    \item \textbf{Short documents} ($<$100 sentences): $\sim$\$0.0002--0.0005 per document
    \item \textbf{Long documents} (400+ sentences, $\sim$15k tokens): $\sim$\$0.002--0.005 per document
\end{itemize}
For a corpus of 1,000 documents, total preprocessing cost ranges from \$0.20 (short docs) to \$5.00 (long docs). Note that for large-scale processing, API rate limits may become a bottleneck; costs assume parallelization is feasible.

\section{Discussion}

\textbf{Why IDC Works:} IDC's improvements stem from aligning chunk boundaries with likely information needs. By predicting questions users might ask, IDC creates ``answer-sized'' segments that contain complete, focused content. This contrasts with fixed-length chunking (which arbitrarily fragments information) and coherence-based methods (which optimize for topical consistency but not query relevance).

\textbf{When IDC Excels:} The largest gains occur on long, heterogeneous documents where static segmentation struggles. Technical manuals (Fiori +60\%), academic papers with diverse sections (arXiv +67\%), and multi-document collections (SQuAD 2-doc +5\%) all benefit substantially. In these cases, IDC's dynamic chunk sizing (larger for broad explanations, smaller for specific facts) outperforms uniform approaches.

\textbf{When IDC Ties Baselines:} On well-structured documents like Qasper academic papers, paragraph boundaries naturally align with distinct topics and questions. Here, simple paragraph-based segmentation achieves comparable results. IDC provides no advantage when document structure already reflects likely query boundaries.

\textbf{Limitations:} IDC depends on LLM-generated intents; if the model fails to predict relevant questions, segmentation quality suffers. Some datasets had small sample sizes (n=15), limiting statistical power. Additionally, IDC's offline processing adds indexing time, though this is acceptable for most applications.

\section{Conclusion}

We introduced Intent-Driven Dynamic Chunking (IDC), a novel approach that segments documents based on predicted user intents. By generating likely questions via an LLM and optimizing chunk boundaries through dynamic programming, IDC produces segments aligned with actual information needs. Evaluation across six diverse QA datasets showed that IDC outperformed traditional chunking methods on five datasets, with R@1 improvements ranging from 5\% to 67\%, while producing 40--60\% fewer chunks with higher answer coverage.

IDC is particularly effective for long, heterogeneous documents where static segmentation fails to isolate relevant content. The approach adds minimal computational overhead suitable for offline indexing, with no impact on query-time latency.

Future work includes extending IDC to multi-hop queries requiring information synthesis across chunks, incorporating real user feedback for adaptive re-segmentation, and exploring domain-specialized intent generation for technical corpora.

\end{document}